\documentclass{article}
\usepackage{amssymb}        % For \checkmark
\usepackage{pifont}         % For \ding{55}
\usepackage{multirow}       % For \multirow
\usepackage[table]{xcolor}  % For \cellcolor
\usepackage{url}
\usepackage{PRIMEarxiv}

\usepackage{microtype}
\usepackage{graphicx}
\usepackage{subfigure}
\usepackage{pifont}
\usepackage{booktabs} % for professional tables
\usepackage{amsmath} % for equations
\usepackage{multirow}
\usepackage{tcolorbox}
\usepackage{textcomp}
\usepackage[table]{xcolor} % for cell coloring
\usepackage{colortbl}      % more table color tools
\AtBeginDocument{%
  }

\begin{document}

\title{Fine-tuning Small Language Models as Efficient Enterprise Search Relevance Labelers}

\author{
Yue Kang$^{1}$\thanks{Both authors have equal contribution.}$\; \, ^\dag$ \quad Zhuoyi Huang$^{1}$\footnotemark[1] \quad Benji Schussheim$^{1}$ \quad Diana Licon$^{1}$ \quad \text{Dina Atia}$^{1}$ \\
\textbf{Shixing Cao}$^{1}$ \quad \textbf{Jacob Danovitch}$^{1}$\thanks{Corresponding authors: \texttt{\{yuekang,jdanovitch,tianweichen,svesal,sosrini\}@microsoft.com}} \quad \textbf{Kunho Kim}$^{1}$ \quad \textbf{Billy Norcilien}$^{1}$ \quad \textbf{Jonah Karpman}$^{1}$ \\ 
\textbf{Mahmound Sayed}$^{1}$ \quad \textbf{Mike Taylor}$^{1}$ \quad \textbf{Tao Sun}$^{1}$ \quad \textbf{Pavel Metrikov}$^{1}$ \quad \textbf{Vipul Agarwal}$^{2}$\thanks{This work was conducted while employed at Microsoft.}  \\
 \textbf{Chris Quirk}$^{1}$ \quad
  \textbf{Ye-Yi Wang}$^{1}$ \quad \textbf{Nick Craswell}$^{1}$ \quad \textbf{Irene Shaffer}$^{1}$ \quad \textbf{Tianwei Chen}$^{1}$\footnotemark[2] \\
   \textbf{Sulaiman Vesal}$^{1}$\footnotemark[2] \quad \textbf{Soundar Srinivasan}$^{1}$\footnotemark[2] \\ 
$^1$Microsoft \quad \; \; $^2$Amazon
}
\maketitle
\begin{abstract}
In enterprise search, building high-quality datasets at scale remains a central challenge due to the difficulty of acquiring labeled data. To resolve this challenge, we propose an efficient approach to fine-tune small language models (SLMs) for accurate relevance labeling, enabling high-throughput, domain-specific labeling comparable or even better in quality to that of state-of-the-art large language models (LLMs). To overcome the lack of high-quality and accessible datasets in the enterprise domain, our method leverages on synthetic data generation. Specifically, we employ an LLM to synthesize realistic enterprise queries from a seed document, apply BM25 to retrieve hard negatives, and use an a teacher LLM to assign relevance scores. The resulting dataset is then distilled into an SLM, producing a compact relevance labeler. We evaluate our approach on a high-quality benchmark consisting of 923 enterprise query–document pairs annotated by trained human annotators, and show that the distilled SLM achieves agreement with human judgments on par with or better than the teacher LLM. Furthermore, our fine-tuned labeler substantially improves throughput, achieving a $17\times$ increase while also being $19\times$ more cost-effective. This approach enables scalable and cost-effective relevance labeling for enterprise-scale retrieval applications, supporting rapid offline evaluation and iteration in real-world settings.
\end{abstract}

\section{Introduction}
\label{intro}
As AI-driven search technology advances, Large Language Models (LLMs) have demonstrated remarkable capabilities in search and relevance labeling~\cite{zhu2024largelanguagemodelsinformation,wang2023surveyfactualitylargelanguage}. However, enterprise search presents unique challenges that go beyond those of traditional web search~\cite{hawking2004challenges}. 
% Enterprise users issue queries over heterogeneous corpora, such as emails, chats, documents, and organizational knowledge bases, that are both keyword-based (e.g., searching by file name or author) and semantic (e.g., asking natural language questions about documents). 
Enterprise search refers to retrieving information from an organization’s heterogeneous internal repositories, such as emails, chats, documents, and knowledge bases, where users issue both keyword-based queries (e.g., searching by file name or author) and semantic queries (e.g., asking natural language questions about documents).
Unlike web search queries, enterprise queries are often ambiguous, persona-specific, and heavily context-dependent, requiring systems to interpret user intent while accounting for content, metadata, and both personal and organizational context~\cite{radlinski2017theoretical}. For example, while the query ``Juno release date'' has a clear intent in web search relating to the 2007 film \footnote{\url{https://en.wikipedia.org/wiki/Juno_(film)}}, the same query in enterprise could have multiple intents, e.g. looking for files containing the release date of a project named Juno or for emails from a colleague named Juno. The differences between enterprise search queries, the focus of our work, versus web search queries are also presented in Table~\ref{tab:query}.

Consequently, labeling relevance in enterprise search is inherently more complex than in web search. Moreover, there is a lack of publicly available, high-quality enterprise search datasets that align with real-world data structures and distributions, limiting the development of enterprise search methods. In contrast, web search has benefited from large-scale public benchmarks such as MS MARCO~\cite{nguyen2016ms} and TREC-CAsT~\cite{dalton2020trec}, which have driven rapid progress in retrieval and relevance labeling. In addition, personalization and contextualization are essential, as relevance depends on both textual metadata and user networks. Finally, enterprise queries are often underspecified, domain-specific, and long-tail, making ranking evaluation particularly challenging due to sparse user engagement and limited human-labeled data. These challenges underscore the value of LLMs for relevance labeling~\cite{choi2024rradistill}. However, while LLM-based judgments~\cite{thomas2024large} have become a standard approach for enterprise relevance labeling, they come with significant drawbacks: LLMs are computationally expensive and have limited throughput, making them inefficient for large-scale labeling tasks. To address these limitations, Small Language Models (SLMs)~\cite{abdin2024phi3technicalreporthighly} offer a promising alternative. With domain-specific fine-tuning on curated synthetic data, SLMs can achieve decent performance comparable to LLMs while significantly improving throughput and reducing costs.

Recent studies have investigated training SLMs for labeling web search queries~\cite{fitte2025augmented}, but enterprise scenarios demand broader coverage of hybrid queries that combine lexical and semantic signals, a setting that remains largely unexplored. To address this gap, we propose a novel synthetic data generation pipeline that fine-tunes SLMs to label enterprise search queries at scale. Our approach leverages enterprise documents to seed the generation of realistic user queries, applies BM25~\cite{robertson1995okapi} to retrieve challenging negatives, and employs LLM-based judgments to create graded relevance labels. These synthetic datasets are then distilled into an SLM, yielding an efficient and cost-effective relevance labeler capable of handling diverse enterprise query types. This method enhances the offline evaluation process for ranking models, offering a scalable alternative to conventional LLM-based labelers and human judges. Our main contributions are as follows:
\begin{itemize}
    \item We propose a novel synthetic data generation pipeline that leverages BM25~\cite{robertson1995okapi} and LLM-based labeling to create high-quality, domain-specific, and realistic training data for enterprise search relevance labeling, addressing existing dataset limitations.
    \item We showcase the effective fine-tuning of an SLM, Phi-3.5 Mini Instruct~\cite{abdin2024phi}, achieving performance comparable to its teacher LLM (GPT-4o) while significantly improving throughput and reducing costs.
    % \item We demonstrate the efficiency of the fine-tuned SLM at production scale, enabling cost-effective offline ranking evaluation and supporting broader information retrieval tasks as a promising alternative to traditional LLM-based or human labeling.
\end{itemize}
These facts demonstrate that the fine-tuned SLM enables cost-effective, production-scale offline ranking evaluation, offering a practical alternative to LLM-based or human labeling.

\section{Related Work}
\label{relatedwork}
\paragraph{Large and Small Language Models}
The rapid progress of generative language models has transformed natural language processing (NLP) and reshaped modern information retrieval (IR), enabling systems to perform advanced reasoning and comprehension across a wide spectrum of applications~\cite{naveed2025comprehensive,wu2024bloomberggpt,thirunavukarasu2023large}. Early innovations such as ELMo~\cite{peters-etal-2018-deep} and BERT~\cite{devlin2019bert} demonstrated that large-scale pretraining on raw text could capture rich contextual dependencies, and later refinements like ELECTRA~\cite{clark2020electra} and DeBERTa~\cite{he2020deberta} introduced more efficient and accurate pretraining strategies. The introduction of sequence-to-sequence architectures, such as T5~\cite{raffel2020exploring}, has enabled a unified treatment of diverse IR tasks, ranging from query rewriting to passage ranking. Embedding-based retrieval methods, such as ColBERT~\cite{khattab2020colbert}, further highlighted how encoders can surpass lexical matchers by capturing nuanced semantic signals. In parallel, decoder-only models such as GPT~\cite{brown2020language,hurst2024gpt} marked the advent of large-scale generative systems and rapidly extended applications to document ranking, conversational search, and summarization~\cite{zhu2023large}, ultimately establishing themselves as state-of-the-art engines for modern IR.

More recently, attention has shifted from larger models to compact yet competitive small language models (SLMs). Families such as Gemma~\cite{team2024gemma}, Llama~\cite{grattafiori2024llama}, and Phi series~\cite{abdin2024phi3technicalreporthighly,abdin2024phi} demonstrate that models with a few billion parameters can offer strong generalization and reasoning capacity while drastically lowering inference cost. While SLMs typically fall short of frontier LLMs in absolute performance, their speed, lower memory footprint, and reduced inference cost make them highly appealing for retrieval and labeling tasks where throughput and scalability are critical~\cite{samarinas2025distillation}. This trade-off, accepting slightly lower peak accuracy in exchange for efficiency, allows SLMs to serve as practical complements to LLMs, enabling high-volume offline evaluation and rapid experimentation without the burden of massive GPU resources.

\paragraph{Relevance Labeling with Language Models}
Relevance assessment has long been a cornerstone of IR, traditionally dependent on human annotations along with heuristic models such as BM25~\cite{robertson1995okapi}. While accurate, these methods were costly and limited in scale. More recently, LLMs have been explored as alternatives to replicate or augment human judgments. \cite{macavaney2023one} introduced one-shot labeling for automatic relevance estimation, showing that LLMs could generate labels efficiently. \cite{abbasiantaeb2024can} examined the use of LLMs to fill relevance judgment holes, and later~\cite{thomas2024large} showed that LLMs could accurately predict searcher preferences, aligning closely with user behavior, while \cite{farzi2025criteria} proposed criteria-based LLM judgments to improve label consistency across dimensions such as topicality and coverage. 

At the same time, fine-tuning techniques such as LoRA~\cite{hu2021loralowrankadaptationlarge} have made it feasible to adapt LLMs to domain-specific ranking tasks~\cite{mehrdad2024large}. More efficient distillation-based methods have also emerged:  \cite{fitte2025augmented} demonstrated that fine-tuned small LLMs could improve the quality of augmented relevance datasets, highlighting the promise of post-training SLMs for scalable evaluation. RRADistill~\cite{choi2024rradistill} distilled large models into SLMs for re-ranking long-tail queries, while Rank1~\cite{weller2025rank1} leveraged reasoning-based LLMs (e.g., OpenAI’s o1~\cite{jaech2024openai}, DeepSeek’s R1~\cite{guo2025deepseek}) and MS MARCO-derived traces to train smaller, explainable, high-performing re-rankers. Collectively, these approaches highlight the promise of synthetic data pipelines and distillation for building compact yet effective retrieval models.   

Despite this progress, the exploration of SLMs for relevance labeling has been largely limited to open-domain, semantically driven retrieval tasks~\cite{fitte2025augmented}. While these works showed that compact models can enhance dataset quality or approximate semantic relevance, enterprise search poses distinct challenges. Queries in enterprise environments often combine sparse keyword matches with semantic intent, and document collections span heterogeneous domains such as Teams chats, emails, and files. Current industrial practice continues to rely on frontier-scale LLMs, combined with careful prompt engineering and chain-of-thought reasoning~\cite{wei2022chain}, to handle this complexity~\cite{thomas2024large}. To the best of our knowledge, no prior work has examined whether fine-tuned SLMs can deliver competitive and resource-efficient labels in this setting. Our work addresses this gap by adapting Phi-3.5-mini for enterprise relevance labeling, demonstrating that carefully curated training datasets and distillation pipelines can make SLMs a faster, more efficient, and practical alternative for large-scale experimentation.  

Synthetic data generation with LLMs to create queries has proven effective for model training in low-resource settings ~\cite{braga2024synthetic}. Retrieval approaches fine-tuned solely on synthetic data have outperformed strong baselines like BM25 and other self-supervised dense retrieval methods across various domain-specific IR tasks ~\cite{inpars}. Approaches such as Promptagator ~\cite{dai2022promptagatorfewshotdenseretrieval} and InPars v2 ~\cite{inparsv2} have enabled the development of open-source rankers that achieve state-of-the-art results on public benchmarks, while DUQGen~\cite{DUQGen} explored unsupervised domain adaptation through clustering and probabilistic sampling to diversify synthetic queries.

% Language model fine-tuning techniques, such as LoRA ~\cite{hu2021loralowrankadaptationlarge}, have been employed to adapt LLMs for domain-specific ranking tasks ~\cite{mehrdad2024large}. Meanwhile, advances in efficient label generation have led to approaches like RRADistill ~\cite{choi2024rradistill}, which distills large models into small language models (SLMs) for re-ranking long-tail queries. More recently, Rank1 ~\cite{weller2025rank1} uses reasoning-based LLMs (e.g., OpenAI’s o1, DeepSeek’s R1) and MS MARCO-derived reasoning traces to train smaller, explainable, and high-performing re-ranking models.

Building on these innovations, we focus on the practical deployment of a Small Language Model (SLM) labeler for enterprise search and retrieval. Specifically, we fine-tune Phi-3.5 Mini~\cite{abdin2024phi3technicalreporthighly} using domain-specific data generated through an LLM-driven query and label pipeline, customized via instruction tuning on synthetic enterprise-style queries and documents. We selected Phi-3.5 Mini and GPT-4o as benchmark models, given their proven performance in recent retrieval and reasoning studies~\cite{zhu2023large}, and because they represent strong exemplars of their respective model classes (compact SLMs vs. frontier-scale LLMs). While we did not include the results with newer models such as GPT-5 or alternative SLM families, we believe our findings to generalize, since the core advantage comes from the data generation and distillation pipeline rather than any model-specific property. Our resulting SLM labeler achieves performance comparable to frontier LLMs while delivering substantial improvements in efficiency and scalability, making it well-suited for large-scale enterprise relevance evaluation.
\begin{table}[t]
    \centering
    \footnotesize% Change the font size here
    \begin{tabular}{|p{2.6cm}|p{5.4cm}|p{5.4cm}|}

    \hline
 \textbf{Query Type}        & \textbf{Enterprise  }                          & \textbf{Web}                                   \\ \hline
    Description & Queries leverage context and internal entities along with semantic understanding & Queries rely on common knowledge and general semantics         \\ \hline 
    Example     & iris project date& structure of the iris \\ \hline
    Key Difference & Further require enterprise-specific knowledge (e.g., “iris” = colleague/project) & Use global/common knowledge (e.g., “iris” = part of the eye) \\ 
    \hline
    \end{tabular}
        \caption{Enterprise vs. web search: both can be semantic, but enterprise queries depend on domain-specific knowledge.}
        \label{tab:query}
\end{table}
Notably, traditional approaches to search relevance labeling, such as manual annotation or direct reliance on frontier LLMs, present significant challenges. Manual annotation is often infeasible in real-world enterprise search due to strict privacy policies and the substantial effort required to go through large-scale datasets. LLMs-based labeling, while capable of producing high-quality labels, is computationally expensive and slow, making it unsuitable for large-scale deployment.

\subsection{Problem Definition}
Given a user query $q$, the objective is to determine the relevance of a set of documents $D = \{d_1, d_2, ..., d_n\}$ to the query $q$. This can be formalized as learning a relevance function $r(q, d_i)$ that assigns a relevance score to each query-document pair, $(q, d_i)$, where $d_i \in D$. We define the problem as follows:

\textbf{Input:}
\begin{itemize}
    \item A query $q \in \mathcal{Q}$, where $\mathcal{Q}$ is the space consisting of all user enterprise queries.
    \item A set of documents $D = \{d_1, d_2, ..., d_n\}$, where $d_i \in \mathcal{D}$ and $\mathcal{D}$ are the space of documents.
\end{itemize}

\textbf{Output:}
\begin{itemize}
    \item A relevance score $r(q, d_i) \in \mathcal{R}$ for each $(q, d_i)$ pair, where $\mathcal{R}$ is the space of relevance scores. In this paper, we adopt an ordinal 0–4 scale, with 0 indicating irrelevance and 4 indicating high relevance. Scores $r(q, d_i)$ are designed to approximate human judgments and align with the quality of state-of-the-art LLMs such as GPT-4o.
\end{itemize}
Our goal is to develop a practical and scalable approach for enterprise search relevance labeling, enabling rapid and cost-effective evaluation of various IR models while ensuring high-quality and consistent supervision in real-world information filtering systems.

\subsection{Datasets}
Table \ref{tab:datasets} provides detailed descriptions of the datasets and specifies their roles across the different phases from model training to evaluation. Specifically, we use four datasets for training and one benchmark dataset for final evaluation. We obtain 1,500 proprietary enterprise documents curated under an eyes-on review setting for synthetic data generation and fine-tuning, while we believe a similar dataset could be obtained from any document platform. Since no public enterprise search dataset is available for evaluation, we curate an internal benchmark consisting of 923 high-quality enterprise query–document pairs annotated by trained human annotators.

\begin{table*}[t]
    \centering
    \renewcommand{\arraystretch}{1.3}
    \scriptsize % Change the font size here
    \begin{tabular}{|p{2.4cm}|p{2cm}|p{5.8cm}|p{5cm}|}
        \hline
        \textbf{Dataset} & \textbf{\# Samples} & \textbf{Description} & \textbf{Usage} \\ 
        \hline
        Eyes-on enterprise documents
        & 1,500 & 
        A proprietary collection of internal documents with associated metadata, curated under an eyes-on review setting. & 
        \textbf{Supervised Fine-tuning:} Used to generate synthetic query–doc–label triplets and fine-tune the model on them. \\ 
        \hline
        INTERS \cite{zhu2024inters} & $\sim$250,000 samples on 20 different tasks & 
        Multi-task tuning dataset covering multiple tasks, including query, document, and query-document relationship understanding. & 
        \textbf{Supervised Fine-tuning:} Used for multi-task tuning. (Green box in Figure \ref{fig:enter-label}) \\ 
        \hline
        TREC-CAsT \cite{dalton2020trec} & 4,000 & 
        Public dataset for Conversational Information Seeking (CIS) with passage-query-human label triplets. & 
        \textbf{Supervised Fine-tuning:} Used $\sim$4k samples for instruct tuning. (Blue box in Figure \ref{fig:enter-label}) \\ 
        \hline
        MS MARCO Passage \cite{nguyen2016ms} & Over 400,000 passages & 
        Open dataset for information retrieval, containing passage-query pairs and some of them have human annotations of scale 0-3. & 
        \textbf{Supervised Fine-tuning:} Used GPT-4o to generate synthetic queries for instruct tuning. (Purple box in Figure \ref{fig:enter-label}) \\ 
        \hline
        Human-labeled enterprise search dataset & 923 & 
        A proprietary dataset of enterprise query–document pairs, annotated by trained human labelers for relevance. & 
        \textbf{Evaluation:} Used as the gold-standard evaluation dataset for final comparison. \\ 
        \hline
    \end{tabular}
    \caption{Datasets used for fine-tuning and evaluation}
    \label{tab:datasets}
\end{table*}

\subsection{Evaluation Metrics}
To assess the effectiveness of our SLM relevance labelers against baseline labelers, we employ two complementary evaluation metrics: Normalized Discounted Cumulative Gain (NDCG)~\cite{wang2013theoretical} and Pairwise Accuracy. NDCG is a widely adopted metric in information retrieval that evaluates the quality of ranked lists with respect to graded relevance, while Pairwise Accuracy measures how well the model preserves the correct relative ordering of documents under the same query. A higher level of both metrics indicates that the model’s predictions are more consistent with the intended relevance ordering. After computing both metrics at the query level, we report their average values across different queries.

\textbf{1. Normalized Discounted Cumulative Gain (NDCG): } We adopt full Normalized Discounted Cumulative Gain (NDCG) to capture model performance across entire rankings under each query. NDCG measures ranking quality by assigning higher gains to more relevant documents and discounting them logarithmically by their rank positions, ensuring that placing relevant items earlier in the list is rewarded more. Unlike truncated variants (e.g., NDCG@k), full NDCG accounts for all retrieved documents for each query, offering a comprehensive and widely recognized evaluation of ranking effectiveness.

\textbf{2. Pairwise Accuracy Metric (Accuracy):} This metric measures how consistently the model preserves the correct relative ordering of documents compared to the ground-truth labels. For each query, we exhaustively consider all possible pairs of documents $(A,B)$ and check whether the model’s prediction of their relative relevance agrees with the ground truth. For example, if both the model and the ground truth indicate that $A$ is more relevant than $B$ ($A > B$), then this pair is counted as correctly aligned. We compute the proportion of correctly aligned pairs across all possible combinations for a query, resulting in a score between 0 and 1, and then average these scores across queries. Table~\ref{tab:model_baseline_comparison_accuracy} illustrates the full computation matrix. A cell value of 1 indicates perfect agreement between the model and the ground truth for that comparison, while 0 indicates a mismatch. 
\begin{table}[ht]
    \centering
    \renewcommand{\arraystretch}{1.3}
    \scriptsize % Change the font size here
    \begin{tabular}{|l|c|c|c|}
        \hline
        \textbf{Model Output} & \textbf{$A \textless  B$} & \textbf{$A = B$} & \textbf{$A \textgreater  B$} \\ 
        \hline
        \textbf{Model says $A \textless  B$} & 1 & 0 & 0 \\ 
        \hline
        \textbf{Model says $A = B$} & 0 & 1 & 0 \\ 
        \hline
        \textbf{Model says $A \textgreater  B$} & 0 & 0 & 1 \\ 
        \hline
    \end{tabular}
    \caption{Accuracy metric details. Here: 1: The model's prediction exactly matches the baseline label, 0: The model's prediction does not match the baseline label.}
    \label{tab:model_baseline_comparison_accuracy}
\end{table}

\textbf{3. Request Per Minute (RPM):} We also report RPM, which is a metric used to measure the rate at which our SLM labeler processes requests on declared hardware. It is commonly used to monitor system performance, rate limits, and scalability. 
{
\begin{equation}
    \text{RPM} = \frac{\text{Number of total requests for labeling}}{\text{Time elapsed for the labeling in minutes}} \nonumber
\end{equation}
}

\section{Methods}\label{methods}
% Although we have access to a large set of real-world user query-document pairs, they are restricted to the eyes-off workspace due to privacy concerns, making their direct use significantly challenging. Additionally, the eyes-off environment imposes further constraints on limited computing resources, restricting large-scale model training and experimentation. Furthermore, click-based interactions within query-document pairs are often strongly biased, as they primarily reflect user engagement rather than true relevance. Users tend to click on higher-ranked results regardless of actual relevance, leading to skewed training data that can reinforce suboptimal ranking behaviors. These challenges necessitate an alternative approach that enables effective training while respecting privacy constraints and mitigating biases in query-document interactions. Therefore, we proposed to generate high-quality synthetic queries with accurate labels based on curated documents.

Unlike web search, where large-scale open-source datasets such as MS MARCO~\cite{nguyen2016ms} and TREC-CAsT~\cite{dalton2020trec} provide high-quality human-labeled query–document pairs, there is no publicly available dataset for enterprise search that captures its unique characteristics. While some enterprise systems may collect query logs, this is not always the case, and oftentimes these contain sensitive, confidential user information which makes it untenable to collect human relevance judgements. Moreover, interaction data such as click logs are inherently biased, as they primarily reflect user engagement rather than true relevance~\cite{vardasbi2020inverse}. For example, users often click on higher-ranked results regardless of their quality~\cite{craswell2008experimental}. These limitations underscore the need for an alternative approach that enables effective training while addressing the absence of feasible training data.

To this end, we propose two approaches for constructing fine-tuning datasets for enterprise relevance labeling without relying on large-scale query logs. Instead, both methods require only a curated set of seed documents. As a result, our methodology is fully automatic and inherently privacy-preserving, making it well suited for enterprise settings where user queries are sensitive, scarce, or unavailable. Specifically, both approaches are based on the idea of generating synthetic queries from the aforementioned seed documents. To generate synthetic queries for fine-tuning our SLM labeler, we investigate two different approaches. The first approach fine-tunes an auxiliary SLM to generate queries given a seed document and a predefined relevance score, enabling synthetic query creation through random sampling of document–score pairs. These generated queries are then used for training the SLM labeler. The second approach, in contrast, involves creating synthetic query-document-score triplets using LLM-based query generation, BM25-based negative document mining, and LLM-assisted labeling, followed by fine-tuning the SLM labeler on these triplets. In practice, we found that the first method struggled to produce high-quality and diverse queries, particularly for keyword-based enterprise search, limiting its utility. Consequently, we focus primarily on the second approach, which demonstrated more consistent and reliable performance. We describe both methods in detail below.

\subsection{Fine-tune SLMs as Query Generators}\label{subsec:querygenerator}
We first fine-tuned the Phi-3.5 Mini Instruct model as a query generator, as this approach enables faster large-scale query generation with controlled relevance levels while reducing the reliance on expensive LLM inference. To the best of our knowledge, there is no prior work on fine-tuning an SLM for enterprise search query generation. Additionally, we previously had fine-tuned the same SLM for semantic web query generation, conditioning it on both the context and a predefined relevance score. This method showed promising results in key evaluation metrics, which motivated us to extend the method to the enterprise setting. The key differences between enterprise search queries, which are the focus of this work, and web search queries are presented in Section~\ref{intro}.
\subsubsection{Experimental Setup}
\label{Experimental&Setup}
\begin{figure*}[!t]
    \centering
    \includegraphics[width=0.93\linewidth]{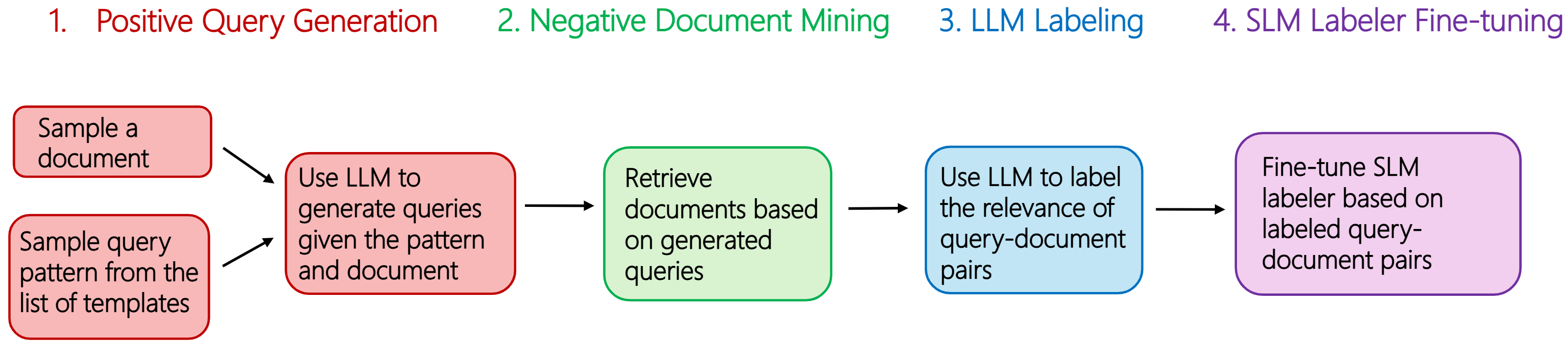}
    \caption{Overall model architecture for our proposed solution.}
    \label{fig:synthetic_pipeline}
\end{figure*}

To generate synthetic queries for fine-tuning our SLM query generator, we utilized GPT-4o to produce both positive and negative enterprise search queries using a carefully designed prompt with explicit instructions for diverse example generation. We then fine-tuned Phi-3.5 Mini Instruct on this synthetic dataset under multiple training configurations aimed at enabling the SLM to generate enterprise-style queries conditioned on document metadata and target relevance levels. However, when we evaluated the fine-tuned model by comparing its outputs against GPT-4o gold-standard judgments (LLM-as-a-judge), we observed that it achieved only 60.8\% binary relevance accuracy, i.e. only 60.8\% of its generated queries matched the intended relevance class (relevant or irrelevant) evaluated by GPT-4o. A deeper manual analysis revealed a strong bias toward generating positive or plausible-sounding queries, even when the prompt explicitly instructed the model to produce irrelevant (negative) ones. As an illustrative instance, we fabricate a sample document to avoid privacy concerns and evaluate our SLM query generator:
\begin{figure*}[t]
    \centering
    \includegraphics[width=0.9\textwidth]{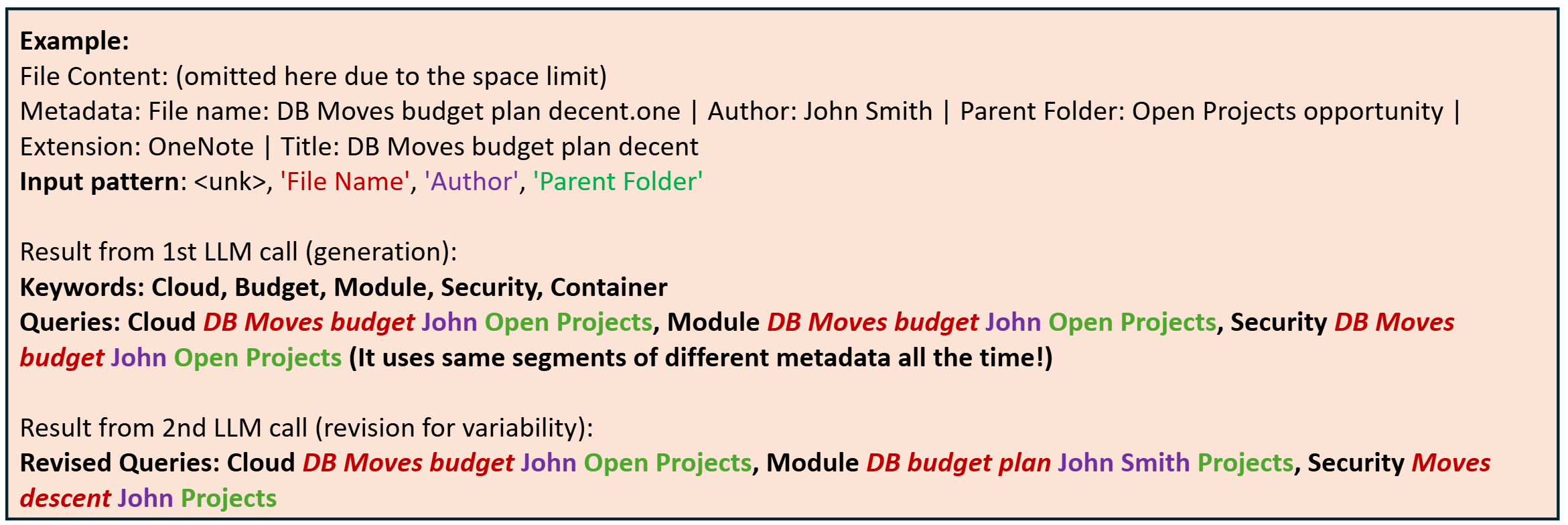}
    \caption{An illustration of our synthetic query generation procedure.}
    \label{fig:twoprompts}
\end{figure*}
\begin{itemize}
    \item Content Summary: A document about how to add a page in Word.
    \item Keywords: page, Word, add
    \item Metadata: File name: "AddPage.docx", Author: Lisa Morrison, Title: "AddPage", File type: "docx", Parent folder: "Word Tutorial"
\end{itemize}

By instructing the model to generate negative queries, we get: "Lisa tutorial Docs", "Add Page" and "Lisa Loop page", but all these three queries are relevant with the document.

We attribute this limitation to two main factors:
\begin{itemize}
    \item Positivity bias in pretraining: Instruction-tuned SLMs like Phi-3.5 Mini are optimized to be helpful and cooperative. When asked to generate a query "for a document that should not be retrieved," the model may still attempt to generate something that sounds relevant or plausible, rather than something truly irrelevant even after fine-tuning.
    \item Limited model capacity:  As a small model, Phi-3.5 Mini may struggle to internalize nuanced instructions like generating “key-word-based queries that appear plausible but are not relevant,” especially in enterprise contexts where relevance boundaries can be subtle.
\end{itemize}

% We believe this limitation arises due to (1) positivity bias in pretraining:  Instruction-tuned SLMs like Phi-3.5 Mini are optimized to be helpful and cooperative. When asked to generate a query "for a document that should not be retrieved," the model may still attempt to generate something that sounds relevant or plausible, rather than something clearly irrelevant even after fine-tuning; (2) limited model capacity: As a small model, Phi-3.5 Mini may struggle to internalize nuanced instructions like generating “key-word-based queries that appear plausible but are not relevant,” especially in enterprise contexts where relevance boundaries can be subtle. {We defer further examples and results to Appendix~\ref{app:querygenerator}.}

Overall, while the fine-tuned model demonstrated a decent ability to generate queries aligned with semantic information from documents, it consistently struggled with the keyword-based nature of enterprise search queries. Unlike traditional web searches, where queries often reflect user intent and semantic understanding, enterprise search further relies on exact keyword matching, file metadata, and structured document attributes. As a result, fine-tuning on SLMs fails to generate meaningful hard negative and diverse queries, and ultimately does not work well under real-world enterprise search applications.

\subsection{Using LLMs and Retrievers for Query Generation}~\label{subsec:llmquerygenerator}
This alternative approach constitutes the core methodology of our work, integrating synthetic data generation with information retrieval techniques. Specifically, we first used GPT-4o to generate and refine synthetic enterprise search queries from documents. Next, we applied ranking algorithms such as BM25 to mine plausible negative documents. With the resulting query–document pairs, we then prompted GPT-4o to assign graded relevance labels using a carefully designed labeling prompt. Finally, we fine-tuned Phi-3.5 Mini Instruct on these query–document–label triplets.

The overall workflow of this synthetic query generation pipeline is illustrated in Figure~\ref{fig:synthetic_pipeline}. It consists of four major components: (1) positive query generation, (2) negative document mining, (3) LLM-based labeling, and (4) SLM labeler fine-tuning. We describe each component in detail below.

\newtcolorbox[auto counter, number within=section]{promptbox}[2][]{%
  colback=gray!10,
  colframe=black!70,
  fonttitle=\bfseries,
  title=Box~\thetcbcounter: #2,
  label={#1}
}

\begin{promptbox}[box:positivegen]{Positive Query Generation Prompt}
\textbf{Role} \\
You are an assistant generating \textbf{keywords} and \textbf{user search queries} from file metadata and content. \\[3pt]
\textbf{Input} \\
-- A structured string of file metadata parts. \\
-- Text content of the file. \\
-- Metadata may contain content placeholders. \\[3pt]
\textbf{Task} \\
\textit{Step 1: Keyword Extraction} \\
-- Extract up to 6 relevant single-word keywords from the content. \\
-- Exclude stop words. \\
-- If content is empty, return an empty list. \\[3pt]
\textit{Step 2: Query Generation} \\
-- Use the metadata string in exact order to form natural queries. \\
-- Replace each content placeholder with keyword. \\
-- Generate 3 distinct queries without reordering parts. \\[3pt]
\textbf{Examples} \\
-- \textit{Few-shot examples shown here}... \\[3pt]
\textbf{Output Format} \\
\texttt{Keywords: k1, k2, k3, ...} \\
\texttt{Queries: q1, q2, q3}
\end{promptbox}

\paragraph{1. Positive query generation:} 
We began with a document dataset and a query pattern template table. In practice, such templates can be constructed by analyzing how users typically form queries in enterprise search. For example, a pattern like \textless author name\textgreater \textless file name\textgreater might yield the query “Lisa budget report.”

In our work, we derived an internal template table containing a ranked list of common query patterns (e.g., \textless folder name\textgreater \textless document type\textgreater \textless keyword\textgreater) along with their frequencies in actual search traffic. However, even without access to such data, synthetic template tables can be generated by systematically enumerating combinations of document metadata fields and keywords, which we found to produce realistic and diverse query structures. The patterns were constructed via entity resolution of query text, ensuring that each segment corresponded to an entity in the file metadata. In addition, templates may include placeholders for content-derived information, where document keywords extracted from the text can be inserted. As such, they accurately reflect how users typically search for documents in workplace settings.

To ensure our synthetic queries mirrored realistic user behavior, we employed weighted sampling over the query pattern table, selecting each pattern with probability proportional to its observed frequency (with replacement). Thus, more frequent patterns were sampled more often, preserving the distributional properties of real enterprise queries. Each sampled pattern served as a template, into which relevant document metadata (e.g., author, title, file type, parent folder) and extracted keywords were dynamically inserted.

% We began with a document dataset and a query pattern template table which which we extracted from real-world query logs. This table contains a ranked list of common query patterns such as \textless author name\textgreater \textless file name\textgreater and \textless folder name\textgreater\textless document type\textgreater \textless keyword\textgreater—along with their observed frequencies in actual enterprise search traffic. These patterns were generated based on entity resolution of the query text, i.e. checking each segment matched an entity in the file metadata, and they reflect how users typically search for documents in workplace settings. For example, the pattern \textless {\textit{author name}}\textgreater \textless { \textit{file name}}\textgreater might lead to a query like ``{\textit{Lisa}} {\textit{budget report}}". {To ensure that our synthetic queries reflected realistic user behavior, we employed weighted sampling over the query pattern table and sampled each pattern with probability proportional to its observed frequency in the query pattern table at each time with replacement.} More frequent patterns were sampled more often, preserving the distributional properties of actual enterprise queries. Each sampled pattern served as a template, into which relevant document metadata (e.g., author, title, file type, parent folder) and extracted keywords were dynamically inserted. 

Once a pattern-document pair was selected, we leveraged GPT-4o to generate three distinct enterprise queries for that pair.
%, and the well-designed prompt of this step is presented in Appendix~\ref{app:llmquery}. 
Each query generation step also included a lightweight keyword summarization phase, where the LLM extracted key thematic terms from the document to incorporate into the queries. This helped ground the queries generated in the actual content of the document. The structure of the used prompt is given in Box~\ref{box:positivegen}. Due to confidentiality constraints, we cannot disclose the full prompt templates. Instead, we provide a high-level outline to illustrate the general structure. The precise wording is not central to our contribution; rather, our key finding is that fine-tuned small language models can effectively perform enterprise relevance labeling across variations in prompt phrasing.

For more complex patterns, those with more components or rare metadata combinations, we observed that the initial queries sometimes lacked variation. For example, if the sampled query template was long, the LLM tended to repeatedly select the same portion of each component, even when explicitly instructed to vary them. To address this limitation, we added an additional LLM-based query revision stage that rewrote the queries for diversity, fluency, and coverage, thereby producing a richer set of realistic queries while preserving their core semantic relevance. This query revision prompt is outlined in Box~\ref{box:positivere}. To illustrate how the two prompts work in sequence, Figure~\ref{fig:twoprompts} shows a example (with fabricated metadata). The synthetic query generation prompt produces three reasonable queries, but they repeatedly use the same metadata segments, limiting diversity. The refinement step then rewrites these queries under strict rules, yielding more diverse outputs.
%The prompt of this revision step is also deferred to Appendix~\ref{app:llmquery}.

This process resulted in a high-quality positive synthetic query-document pairs pool and yielded an average of three diverse, metadata-aligned queries per document-template pair.

\begin{promptbox}[box:positivere]{Positive Query Revision Prompt}
\textbf{Role} \\
You are an assistant validating and revising \textbf{user-generated search queries}. \\[3pt]
\textbf{Input} \\
-- Metadata string (with possible keyword placeholders) \\
-- Keywords list \\
-- Three generated queries \\[3pt]
\textbf{Task} \\
\textit{Step 1: Validation} \\
-- Ensure queries follow metadata order and structure. \\
-- Replace each content placeholder with keyword. \\
-- Avoid redundancy; keep queries short and natural. \\[3pt]
\textit{Step 2: Modification} \\
-- Revise queries so that each metadata part is phrased differently across the three queries. \\[3pt]
\textbf{Examples} \\
-- \textit{Few-shot examples shown here}... \\[3pt]
\textbf{Output Format} \\
\texttt{Revised Queries: query1, query2, query3}
\end{promptbox}

\paragraph{2. Negative document mining:} To construct a balanced and high-quality training set for relevance modeling, we employ the Okapi BM25 retrieval algorithm~\cite{robertson2009probabilistic} to mine negative documents for each synthetic query. BM25 is a strong lexical-based baseline widely used in traditional information retrieval, and it allows us to simulate a realistic retrieval environment in the absence of human click logs or explicit negative labels.

For each synthetically generated query, we use BM25 to retrieve the top-$k$ documents from the corpus, excluding the original source document from which the query was generated. Importantly, we do not assume that the source document is the only relevant document, nor do we require all retrieved documents to be strictly non-relevant. Instead, the retrieved candidates are treated as plausible retrieval results that may span a range of relevance levels. In other words, our goal at this stage is not to pre-classify retrieved documents as relevant or non-relevant, but to collect a realistic spectrum of relevance levels, with final relevance judgments deferred to the subsequent LLM-based annotation stage.

In practice, BM25 tends to return documents with varying degrees of lexical and semantic overlap with the query. Some retrieved documents may be highly relevant (e.g., discussing the same entity or topic), while others are only partially related or largely irrelevant. This behavior aligns well with our objective to construct training data that covers a diverse spectrum of relevance scores. As a result, the retrieved documents naturally form a mixture of hard negatives, weak positives, and near-miss cases, reflecting the ambiguity commonly observed in real-world enterprise search. Such diversity is crucial for teaching the model to distinguish truly relevant documents from those that are only topically or semantically adjacent. For example, given the query ``Lisa budget report'' (derived from a document authored by Lisa Morrison titled ``2023 Budget Report'' mentioned in Subsection~\ref{subsec:querygenerator}), BM25 may retrieve another document such as “Lisa travel plan” or “2022 Budget Draft”. These results are lexically similar and thus plausible retrievals, but they are not the intended match, making them suitable hard negatives for training.

% These retrieved documents are treated as hard negatives or plausible negatives: they share surface-level lexical similarity with the query but are not the intended match. Documents assigned higher relevance scores tend to exhibit greater lexical and semantic overlap with the query (e.g. relevance score 2-3), whereas this overlap generally diminishes as the relevance score decreases (e.g. relevance score 0-1). This step is crucial for teaching the model to distinguish between truly relevant documents and those that are topically or semantically adjacent. 

In our experiments, we set $k=4$, which empirically yields a uniform distribution of relevance labels across levels 0–4 after LLM-based annotation. This balance is desirable for training a robust relevance labeler, as it prevents over-representation of trivially irrelevant examples while encouraging the model to learn subtle distinctions between closely related documents. In practice, $k$ can be adjusted dynamically based on corpus size and diversity to maintain balanced coverage across relevance levels. For example, if the LLM-based annotation results in an over-representation of higher relevance scores, we can increase $k$ to retrieve additional candidate documents and annotate only the newly introduced query–document pairs, while retaining previously labeled pairs, thereby incrementally rebalancing the dataset without reprocessing the entire corpus.

\paragraph{3. LLM labeling:} The next stage in our data pipeline is LLM-based labeling, where GPT-4o assigns relevance scores to each query–document pair. This step is critical for distilling the retrieval capabilities of LLMs into our SLM labeler. We adopted a generic labeling prompt template, which instructs GPT-4o to output a discrete relevance score, ranging from 0 (irrelevant) to 4 (highly relevant), for each query–document pair, as outlined in Box~\ref{box:label}.

To ensure accuracy and consistency, we introduced an additional quality-control round after the initial labeling. In this phase, we automatically checked GPT-4o outputs for formatting issues and semantic inconsistencies. Out of roughly 24,000 samples, only 15 cases contained invalid or incomplete outputs (e.g., missing scores), which were automatically re-labeled. Furthermore, we implemented a post-labeling filtering mechanism to preserve dataset quality. Specifically, if a synthetically generated positive query from Step 1 received an unexpectedly low score (0 or 1), we re-labeled it with GPT-4o using the same prompt. If the low score persisted, we retained it as a valid negative case; otherwise, we discarded the query. This filtering process prevents the model from learning contradictory signals and enforces a consistent training distribution.

After completing these four steps, we obtained a high-quality set of synthetic query–document–label triplets, which served as the core training dataset for fine-tuning our SLM labeler.

\begin{figure*}
    \centering
    \includegraphics[width=0.96\linewidth]{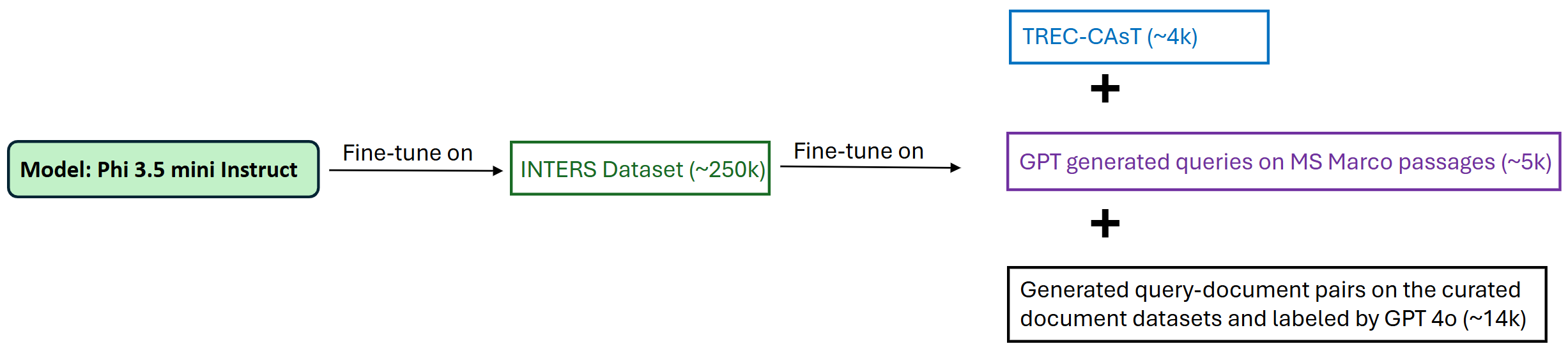}
    \caption{SLM labeler supervised fine-tuning (SFT) workflow: we first fine-tuned Phi-3.5 Mini Instruct on the INTERS dataset for multi-task pre-training, and then fine-tuned on the resulting model again with a combination of synthetic and open-sourced datasets.}
    \label{fig:enter-label}
\end{figure*}

\paragraph{4. SLM labeler fine-tuning:} Finally, we fine-tuned the Phi-3.5 Mini Instruct model using the labeled query-document pairs. To optimize the instruct fine-tuning process, we adapted the same prompt template for LLM labeling in the previous step after removing the instruction for the model to generate explanations alongside the predicted relevance score. Through empirical analysis, we found that including explanations, though useful for interpretability, led to inconsistent training signals and introduced unnecessary verbosity in the trained SLM's output. To further improve the generalization ability of the fine-tuned small language model, we employed multi-task tuning~\cite{zhang2023instruction}. As illustrated in Figure~\ref{fig:enter-label}, this phase incorporated several additional datasets to expose the model to a broader range of query-document relevance scenarios. Specifically, we include:
\begin{itemize}
    \item INTERS~\cite{zhu2024inters}: It consists of data sets for 19 different small tasks about query and document understanding. We used it for multi-task tuning to enhance models' generalizability.
    \item TREC-CAsT~\cite{dalton2020trec}: This dataset consists of semantic web query-passage pairs with human labels on scale of 0-4.
    \item MS MARCO passages~\cite{nguyen2016ms}: It consists of over 400K passages from various sources. We use GPT-4o to generate sematic queries under 0-4 relevance scale under randomly selected passages based on a chain-of-thought idea. 
\end{itemize}
Although our primary goal is enterprise query labeling, we deliberately incorporated semantic web query datasets from the public domain. We believe this broader supervision helps the model develop a stronger understanding of query–document relevance beyond strictly keyword-matching setting, thereby improving both generalization and robustness. Specifically, as shown in Figure~\ref{fig:enter-label}, we first trained the Phi-3.5 Mini Instruct on INTERS for multi-task tuning to enhance its robustness and overall capabilities, and then we continuously fine-tuned the trained model on a combination of three datasets we mentioned above: synthetic enterprise query-document-label triplets, TREC-CAsT with human labels, and MS MARCO passages with synthetic queries. 

We train the model for 2 epochs using a maximum sequence length of 4096. The effective batch size is 32, achieved by setting the per-device batch size to 4 and using 8 gradient accumulation steps. Training logs are recorded every 40 steps. Evaluation is performed every 80 steps using a separate human-labelled test dataset evaluation dataset that does not influence the training process, as there is no early stopping or feedback loop from the evaluation set. Both training and evaluation use a batch size of 4 per device. The fine-tuning was conducted on a cluster of eight A100 GPUs. It achieves a training speed of approximately 5.84 samples per second on the GPU cluster.

\begin{promptbox}[box:label]{Relevance Labeling Prompt}
\textbf{Role} \\
You are an enterprise search quality rater evaluating file/message relevance. \\[6pt]
\textbf{Task} \\
Given a \textbf{query} and an \textbf{entity} (document), assign a relevance score in the range of 0--4: \\ 
-- 4 = ideal quality, should be the ideal result \\
...\\
-- 0 = bad quality, should never be shown \\[6pt]
\textbf{Input} \\
-- Query text \\
-- Document metadata and highlights \\[6pt]
\textbf{Output} \\
\texttt{Score: <between 0 and 4>}
\end{promptbox}

\section{Experimental Results}
\label{exp&results}
In this section, we present a detailed evaluation of the performance of the fine-tuned Phi-3.5 Mini Instruct model compared to the base model (vanilla Phi-3.5 Mini Instruct) and GPT-4o. The results show that fine-tuning significantly enhances the alignment of the Phi-3.5 Mini model with human relevance judgments, matching or even marginally surpassing GPT-4o on both evaluation metrics. 

\begin{figure*}[t]
    \centering
    % First figure
    \begin{minipage}[t]{0.4\textwidth}
        \centering
        \includegraphics[width=\linewidth]{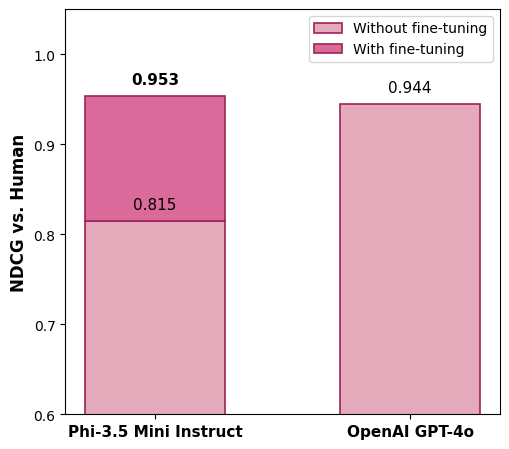}
    \end{minipage}
    \hspace{0.68 cm}
    % Second figure
    \begin{minipage}[t]{0.4\textwidth}
        \centering
        \includegraphics[width=\linewidth]{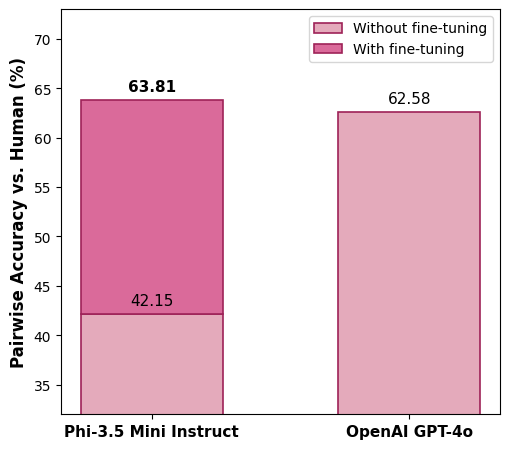}
    \end{minipage}        
    \caption{Performance results comparing the vanilla and fine-tuned SLM models against GPT-4o on two key evaluation metrics for the enterprise search test dataset.}
    \label{fig:performance_results}
\end{figure*}

\paragraph{Fine-tuning Impact:} Fine-tuning has shown a substantial impact on model performance. As highlighted in Figure \ref{fig:performance_results}, the SLM-Human NDCG increased from 0.815 to 0.953, reflecting a significant improvement in the model's alignment with human judgments. Similarly, pairwise accuracy increased from 42.15 to 63.81, surpassing even GPT-4o on both key metrics (NDCG: 0.944, Accuracy: 62.58). This confirms that adapting Phi-3.5 Mini to enterprise relevance tasks yields not only competitive but superior performance, underscoring the effectiveness of our synthetic data generation and fine-tuning pipeline.

We also conducted comprehensive ablation studies on the training data, with results summarized in Table~\ref{tab:performance_results2}. Our key findings are as follows:

%Notably, this fine-tuning outperforms GPT-4o in terms of SLM-Human PAM, which confirms that our method of adapting Phi-3.5 Mini for enterprise relevance tasks yields better results for this application. The fine-tuned model also showed improvements in the SLM-LLM PAM (from 68.16 to 79.66) and accuracy (from 41.38 to 66.16). Although LLM-Human PAM remains consistent at 74.49, fine-tuning improves the model performance in other key metrics, demonstrating the enhanced capability of the Phi-3.5 Mini model compared to the base model. 

\paragraph{Data Size Impact:} It is widely known that more training data may yield better results; however, our experiments show that this effect plateaus beyond a certain point. The fine-tuning results highlight the role of data size in shaping model performance. As shown in Table~\ref{tab:performance_results2}, increasing the amount of synthetic training data from 14K to 24K examples (row 2 vs. row 3) yields only marginal gains. With 14K examples, fine-tuning produced a substantial improvement in SLM-Human NDCG (from 0.815 to 0.953) and pairwise accuracy (from 42.15 to 63.81). Expanding to 24K examples, however, resulted in nearly identical performance (NDCG: 0.954, Accuracy: 63.55). A similar pattern is observed when comparing row 4 and row 5, both without multi-task tuning. This suggests that while scaling data initially drives significant improvements, beyond a certain threshold additional synthetic data offers diminishing returns. 

\paragraph{Multi-task Tuning:} Multi-task tuning demonstrated a notable improvement in aligning model outputs with human relevance judgments. As shown in Table~\ref{tab:performance_results2}, the impact of multi-task tuning becomes clear when comparing row 2 (with multi-task) against row 4 (without). Under the same 14K synthetic training dataset, the model with multi-task tuning achieved NDCG of 0.953 and accuracy of 63.81, compared to 0.946 and 62.41 without multi-task. This increase suggests that multi-task tuning, in conjunction with instruct tuning, effectively enhances the model’s ability to generalize to human relevance preferences. A similar pattern holds when comparing row 3 (with multi-task) and row 5 (without) under the 24K synthetic dataset, where NDCG improves from 0.945 to 0.954 and accuracy from 62.70 to 63.55.

To further validate the role of our synthetic queries, we fine-tuned the model without including the synthetic enterprise query dataset, while keeping all other open-source datasets unchanged (i.e., excluding the data in the black box in Figure~\ref{fig:enter-label}). As shown in row 7 of Table~\ref{tab:performance_results2}, this variant yields only marginal change over the vanilla model, with NDCG moving from 0.815 to 0.826 and Accuracy from 42.16 to 42.88. This outcome aligns with our expectation: existing open-source datasets are primarily designed for web search query understanding and thus do not solely capture the characteristics of enterprise search relevance labeling. These results confirm that our synthetic enterprise queries provide the primary training signal, while multi-task tuning offers an auxiliary benefit by improving robustness via multi-task tuning.

\paragraph{Training Dataset Quality:} As detailed in Subsection~\ref{subsec:querygenerator}, we refine the generated queries from the LLM through a query refinement step to enhance their credibility and diversity. When this refinement was applied after synthetic query generation, the model (row 2 in Table~\ref{tab:performance_results2}) achieved strong performance with NDCG of 0.953 and accuracy of 63.81. In contrast, reusing the same 14K raw synthetic queries without refinement (row 6) led to a measurable drop (NDCG 0.943, Accuracy 60.97). This underscores the critical importance of training data quality. Moreover, increasing the dataset size from 14K to 24K examples (row 2 vs. row 3) produced almost identical results, highlighting that for fine-tuning the SLM, improving data quality has a greater impact than simply adding more data, once the dataset surpasses a reasonably large threshold.

\paragraph{Throughput Analysis:} Our fine-tuned SLM achieved 873.33 RPM on a single A100 GPU, which extrapolates to nearly 7K RPM on a cluster of eight A100s. This throughput is an order of magnitude faster than typical LLM labelers, which are generally limited to the hundreds of RPM range.

\paragraph{Cost Analysis:} Cost is a major consideration, since large-scale evaluation and training data labeling are both frequent and expensive. Our fine-tuned Phi-3.5 Mini Instruct offers a substantial advantage over large language models such as GPT-4o, being nearly 19× cheaper on both input and output tokens. Specifically, the cost is \$0.13 vs. \$2.50 per 1M input tokens and \$0.52 vs. \$10.00 per 1M output tokens~\cite{openaiprice}. This cost efficiency, combined with strong alignment to human judgments, makes trained SLM an attractive option for scalable enterprise relevance labeling.

\paragraph{Statistical Significance Analysis}

To further validate the effectiveness of our fine-tuned labeler, we conducted statistical significance testing against the GPT-4o baseline. Our benchmark dataset consists of 923 query–document pairs with human annotations, spanning 228 distinct queries used for evaluation. For each query, we computed the pairwise accuracy and NDCG of both our fine-tuned labeler and GPT-4o with respect to human annotations. Let $\delta_q = \text{labeler}_q - \text{GPT-4o}_q$ denote the per-query performance difference.

Since our goal is to demonstrate that our labeler is on par with the GPT-4o labeler with significantly less throughput and cost, we performed a one-sided paired non-inferiority test using the Wilcoxon signed-rank test~\cite{wilcoxon1992individual}. The null hypothesis assumes that our model performs worse than the baseline by more than a predefined non-inferiority margin $\Delta$, i.e.,
$$H_0: \mathbb{E}(\delta_q) \leq -\Delta, \quad \quad H_1: \mathbb{E}(\delta_q) > -\Delta.$$
We adopt conservative non-inferiority margins of $\Delta = 0.1\%$ for pairwise accuracy and $\Delta=0.0001$ for NDCG, which represent negligible differences relative to the scale of the respective metrics.

The resulting one-sided Wilcoxon signed-rank tests yield p-values of $0.012$ for pairwise accuracy and $0.00098$ for NDCG, both of which are below the $0.05$ significance level. Therefore, we reject the null hypothesis for both metrics, indicating that our fine-tuned labeler is non-inferior to the GPT-4o baseline within the specified small margins. These results provide statistical evidence, from both pairwise accuracy and NDCG, that our fine-tuned model achieves performance comparable to the strong GPT-4o labeler.

\begin{table*}[t]
    \centering
    \renewcommand{\arraystretch}{1.3}
    \setlength{\tabcolsep}{10pt}
    \resizebox{0.95\textwidth}{!}{\begin{tabular}{c|c|c|c|cccccc}
        \hline\hline
       \multirow{2}{*}{\textbf{Row}}& \multirow{2}{*}{\textbf{Multi-task Tuning}}& \multirow{2}{*}{\textbf{Data Revision}}& \multirow{2}{*}{\textbf{Synthetic Data Size}} & \multicolumn{2}{c}{\textbf{SLM-Human}} & \multicolumn{2}{c}{\textbf{SLM-LLM}} & \multicolumn{2}{c}{\textbf{LLM-Human}}\\
        \cline{5-10}
        & & & &  NDCG & Accuracy (\%) & NDCG & Accuracy (\%) &  NDCG & Accuracy (\%) \\ \hline
          1 &\cellcolor{gray!30}{--}& \cellcolor{gray!30}{--} &\cellcolor{gray!30}{-- (Vanilla SLM)} & 0.815 & 42.15 & 0.820 & 41.38 & {\multirow{6}{*}{0.944}} & \multirow{6}{*}{62.58} \\
           2&\checkmark & \checkmark & 14K  &  {0.953} & 63.81 & 0.951 & 66.16 &  &  \\
           3&\checkmark & \checkmark & 24K  & {0.954} & 63.55 & 0.950 & 66.10 &  & \\
           4&\ding{55} & \checkmark & 14K  & 0.946 & 62.41 & 0.950 & 66.72 &  & \\
           5&\ding{55} & \checkmark & 24K  & 0.943 & 62.50 & 0.948 & 66.83 &  & \\
           6&\ding{55} & \ding{55} & 14K & 0.943 & 60.97 & 0.948 & 66.05 &  & \\
           7&\checkmark & \checkmark & 0 (Only Public Data) & 0.826 & 42.88 & 0.824 & 42.19 &  & \\
        \hline\hline
    \end{tabular}}
    \caption{Performance results comparing the vanilla and trained SLMs under different fine-tuning configurations for ablation studies. The last three column groups present the pairwise metrics based on labels from SLM/SLM/LLM against Human/LLM/Human.}
    \label{tab:performance_results2}
\end{table*}

\paragraph{Qualitative Analysis: } After examining examples from the human-annotated evaluation set, we observe a clear improvement in our fine-tuned labeler over the vanilla SLM. In particular, the original SLM and even GPT-4o tend to avoid assigning the lowest relevance score, even for documents that are completely off-topic, leading to overly optimistic relevance judgments. By contrast, our fine-tuned query labeler demonstrates greater calibration and confidence: it is able to correctly assign a score of 0 when a passage is entirely irrelevant, thereby aligning more closely with human annotations. 

These results collectively demonstrate that carefully fine-tuning SLMs with high-quality synthetic data not only achieves a level of alignment with human judgments comparable to that of large LLMs, but also enables substantially more scalable and cost-efficient relevance labeling.

\section{Conclusion}
Our work presents an efficient and scalable approach to enterprise search relevance labeling by fine-tuning Small Language Models (SLMs) with a synthetic data generation pipeline. The fine-tuned Phi-3.5 Mini Instruct achieves GPT-4o–level accuracy while significantly improving throughput and reducing costs, making large-scale query–document relevance labeling practical. By leveraging GPT-4o for query generation, BM25 for hard negative mining, and LLM-based judgments for labeling, we create high-quality synthetic training data that enables SLMs to perform competitively with LLMs.

Empirical results show our fine-tuned model achieves SLM Human NDCG of 0.953 and pairwise accuracy of 63.81, outperforming GPT-4o (NDCG: 0.944, Accuracy: 62.58). These findings highlight that compact, well-trained SLMs can match or even outperform large models in aligning with human relevance judgments, while enabling faster and more sustainable annotation pipelines. More broadly, this work highlights the potential of synthetic data–driven training to bridge the gap between high-quality human-aligned evaluation and scalable model development, offering a practical path toward deploying lightweight models for data-intensive NLP tasks.
\bibliographystyle{unsrt}  
\bibliography{arxiv}
\end{document}